\documentclass{PoS}
 
\usepackage{subfigure}
\usepackage{mathrsfs}
\usepackage{xspace}

\newcommand {\ie}{\mbox{i.e.}\xspace}     
\newcommand {\eg}{\mbox{e.g.}\xspace}     

\newcommand{\AC}{ATLAS Collaboration}
\newcommand{\CC}{CMS Collaboration}

\title{Recent Progress on 3D Silicon Detectors}

\ShortTitle{Recent Progress on 3D Silicon Detectors}

\author{\speaker{J{\"o}rn Lange}\\
        Institut de F\'{i}sica d'Altes Energies (IFAE),\\
        08193 Bellaterra (Barcelona), Spain\\
        E-mail: \email{jlange@ifae.es}}

\abstract{3D silicon detectors, in which the electrodes penetrate the sensor bulk perpendicular to the surface, have recently undergone a rapid development from R\&D over industrialisation to their first installation in a real high-energy-physics experiment. Since June 2015, the ATLAS Insertable B-Layer is taking first collision data with 3D pixel detectors. At the same time, preparations are advancing to install 3D pixel detectors in forward trackers such as the ATLAS Forward Proton detector or the CMS-TOTEM Proton Precision Spectrometer. For those experiments, the main requirements are a slim edge and the ability to cope with non-uniform irradiation. Both have been shown to be fulfilled by 3D pixel detectors. For the High-Luminosity LHC pixel upgrades of the major experiments, 3D detectors are promising candidates for the innermost pixel layers to cope with harsh radiation environments up to fluences of $2\times10^{16}$\,n$_{eq}$/cm$^2$ thanks to their excellent radiation hardness at low operational voltages and power dissipation as well as moderate temperatures. This paper will give an overview on the recent developments of 3D detectors related to the projects mentioned above and the future plans.}

\FullConference{24th International Workshop on Vertex Detector -VERTEX2015-\\
		1-5 June 2015\\
		Santa Fe, New Mexico, USA}

\begin{document}

\section{Introduction to 3D Detectors}
Segmented silicon detectors are widely used as tracking devices in high-energy-physics (HEP) experiments. However, for cutting-edge present and future applications, there are always increasing demands for higher radiation hardness and reduction of the insensitive area at the edge of the devices.

A modern sensor technology that meets those requirements is the 3D silicon detector technology~\cite{Parker3D}. In contrast to traditional planar sensors, where the electrodes are implanted at the surface, in 3D detectors the electrodes penetrate the sensor bulk as columns vertical to the surface (see Fig.~\ref{fig:FBKCNMdesign} top). This allows to decouple the electrode distance from the sensitive detector thickness. Hence the electrode distance can be tailored and reduced to values as small as $\approx$70\,$\mu$m in today's applications and $\approx$30$\mu$m in the future. This leads to a lower depletion voltage and a smaller drift distance, resulting in faster charge collection and thus less trapping at radiation-induced defects. At the same time, the sensor thickness can be kept at moderate values ($\approx$200\,$\mu$m in today's applications, in the future potentially lower) to allow sufficient charge deposition by ionising particles. As a result, the signal is still high at moderate voltages after heavy irradiation, making this technology very radiation-hard~\cite{bib:Cinzia2009,Koehler,GFradiationHard}. Another advantage is that active~\cite{bib:activeEdge1} or slim~\cite{FBKcutStudy1,FBKcutStudy2,AFP3D} edges are a natural feature of the 3D technology, allowing to highly reduce the insensitive detector edge. However, besides these advantages, the 3D technology comes at the cost of a more complex production process, leading to a longer production time, lower yields and higher costs. Also, the sensor capacitance is typically higher in 3D detectors. Furthermore, there can be a non-uniform response from the 3D columns (which represent dead material) and low-field regions, especially at perpendicular incidence, which can be however reduced by introducing a small tilt of the sensor with respect to the particle incidence.

From proposing the idea in 1997~\cite{Parker3D} to now, 3D detectors have undergone a rapid development from generic Research\&Development (R\&D) to their first applications. Currently we are witnessing the first use of 3D pixel detectors in a HEP experiment, namely in the ATLAS~\cite{ATLAS} Insertable B-Layer (IBL)~\cite{bib:IBL,bib:IBL3Dprod,bib:IBLprototypes}, which has already taken first collision data starting from June 2015. At the same time, the second installations of 3D detectors are imminent, namely in forward trackers such as the ATLAS Forward Proton (AFP) detector~\cite{AFPTDR,AFP3D}, which is planned to be installed in the winter shutdown 2015/16, or the CMS~\cite{CMS}-TOTEM~\cite{TOTEM} Precision Proton Spectrometer (PPS)~\cite{PPSTDR} to be installed in 2016. Simultaneously, R\&D is on-going to develop a new generation of 3D detectors meeting the stringent radiation-hardness requirements of the High-Luminosity(HL)-LHC upgrades~\cite{bib:HLLHC} of ATLAS and CMS, foreseen for 2024.

This paper gives an overview on the present 3D technologies, the status of operation and development of 3D pixel detectors for IBL, AFP and PPS, as well as on the status and plans for the development of 3D detectors for the HL-LHC upgrades.

\section{Overview on Today's 3D Technologies}
\label{sec:technology}

\begin{figure}[bt]
	\centering
	 \includegraphics[width=15cm]{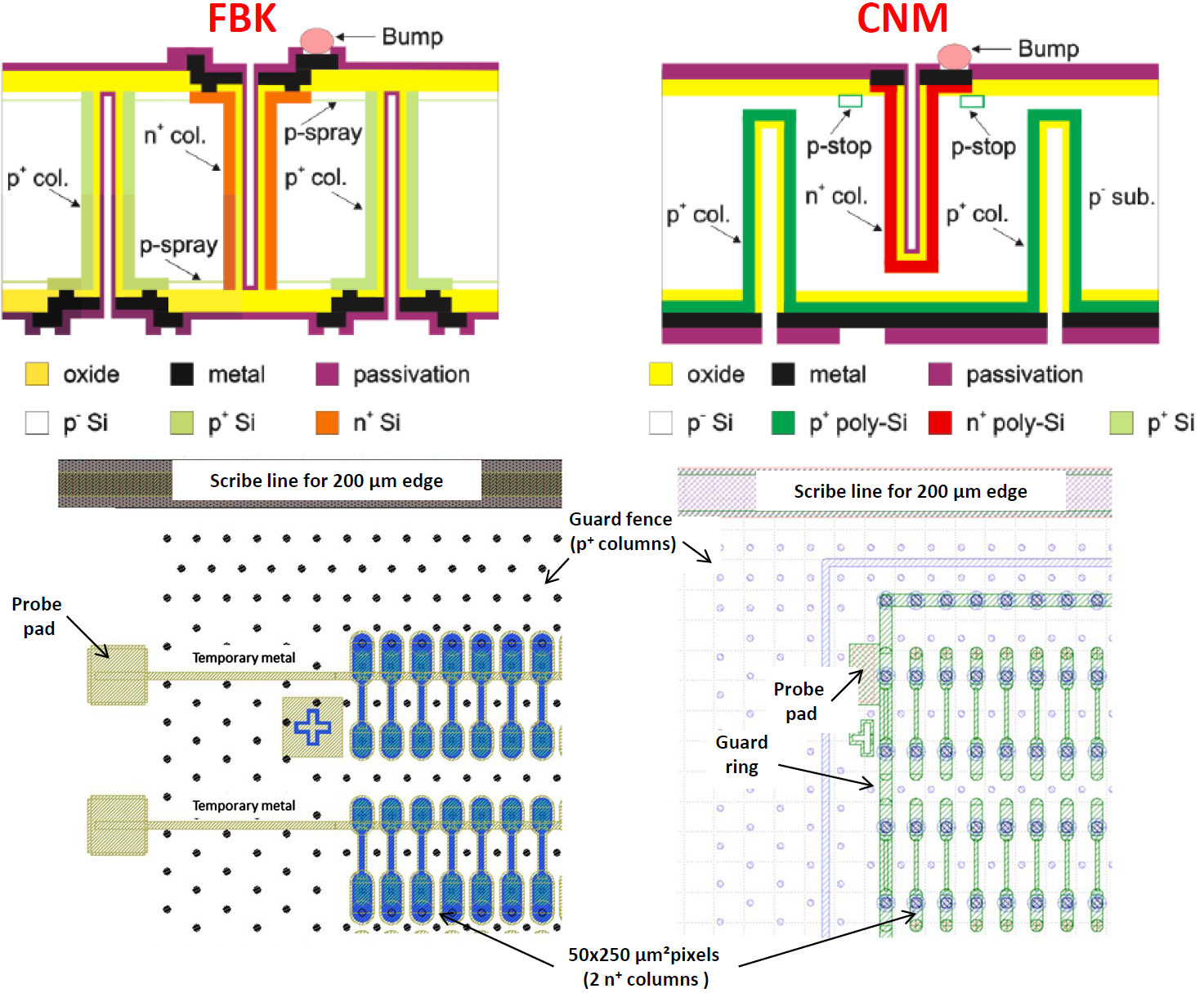}
	\caption{Layout from the side (top) and the top (bottom) of 3D FE-I4 pixel sensors produced with the double-sided technology by FBK (left) and CNM (right) with fully and non-passing-through columns, respectively. Original layout courtesy of FBK and CNM.}
	\label{fig:FBKCNMdesign}
\end{figure}

The most crucial step of the 3D technology is the etching of column-like holes into the silicon bulk\footnote{In the following assumed to be typically p-type due to a better radiation hardness, although in the past also n-type substrates have been used.}. This is typically performed with the deep reactive ion etching (DRIE) technology~\cite{3Detch}. The rows of holes are then alternately n$^+$ and p$^+$ doped to obtain junction and ohmic electrodes, respectively (see Fig.~\ref{fig:FBKCNMdesign}). Two production techniques have been developed: single and double-sided processes. 

In the single-sided technology, which was the originally proposed one, pursued at SNF (Stanford)\footnote{Stanford Nanofabrication Facility, Stanford, CA, United States}~\cite{SNF} and recently also at SINTEF (Oslo)\footnote{SINTEF MiNaLab, Blindern, N-0314 Oslo, Norway}~\cite{sintef}, both column types are etched completely through the sensitive wafer from the front side, requiring a support wafer for mechanical stability, which has to be removed afterwards. Typically the bias voltage needs to be applied at the front side (\ie at the bump-bond side) with the help of an overhanging bias tab. An important advantage of processing on a support wafer is the option to etch the single sensors out of the wafer instead of cutting, followed by heavily doping the sidewalls and hence implementing active edges~\cite{bib:activeEdge1}. This results in only few $\mu$m of insensitive material at the edges.

The double-sided technology was developed and is pursued by CNM (Barcelona)\footnote{Centro Nacional de Microelectronica, CNM-IMB (CSIC), Barcelona, E-08193, Spain}~\cite{CNM1,bib:CNMIBLProduction} and FBK (Trento)\footnote{Fondazione Bruno Kessler, FBK-CMM, Via Sommarive 18, I-38123 Trento, Italy}~\cite{FBK1,bib:FBKIBLProduction}. The n$^+$ and p$^+$ columns are etched from different sides of the sensor, so that a support wafer is not needed and the bias voltage can be applied easily at the back side as for planar sensors. Hence, the process and assembly complexity are reduced. For the IBL production (see Sec.~\ref{sec:IBL}), FBK uses completely-passing-through columns, whereas CNM produces non-passing-through columns that stop about 20\,$\mu$m before reaching the other side as sketched in Fig.~\ref{fig:FBKCNMdesign} (top). For inter-channel isolation, FBK uses the p-spray and CNM the p-stop technique. Without the use of a support wafer, no fully active edges can be produced. However, the edge-termination solutions of a guard fence of p$^+$ columns (FBK) or a guard fence plus a 3D guard ring (CNM) also allow very slim insensitive edges of only about 10\,$\mu$m or 150\,$\mu$m, respectively~\cite{FBKcutStudy2,AFP3D} (see Sec.~\ref{sec:Forward}). Another difference is the on-wafer-selection method using current-voltage characteristics, which is performed on temporary metal layers at FBK and on the 3D guard ring at CNM (see Sec.~\ref{sec:IBL} for further discussion).

\section{Charge Collection Summary}
\label{sec:CC}

\begin{figure}[bt]
	\centering
	 \includegraphics[width=10cm]{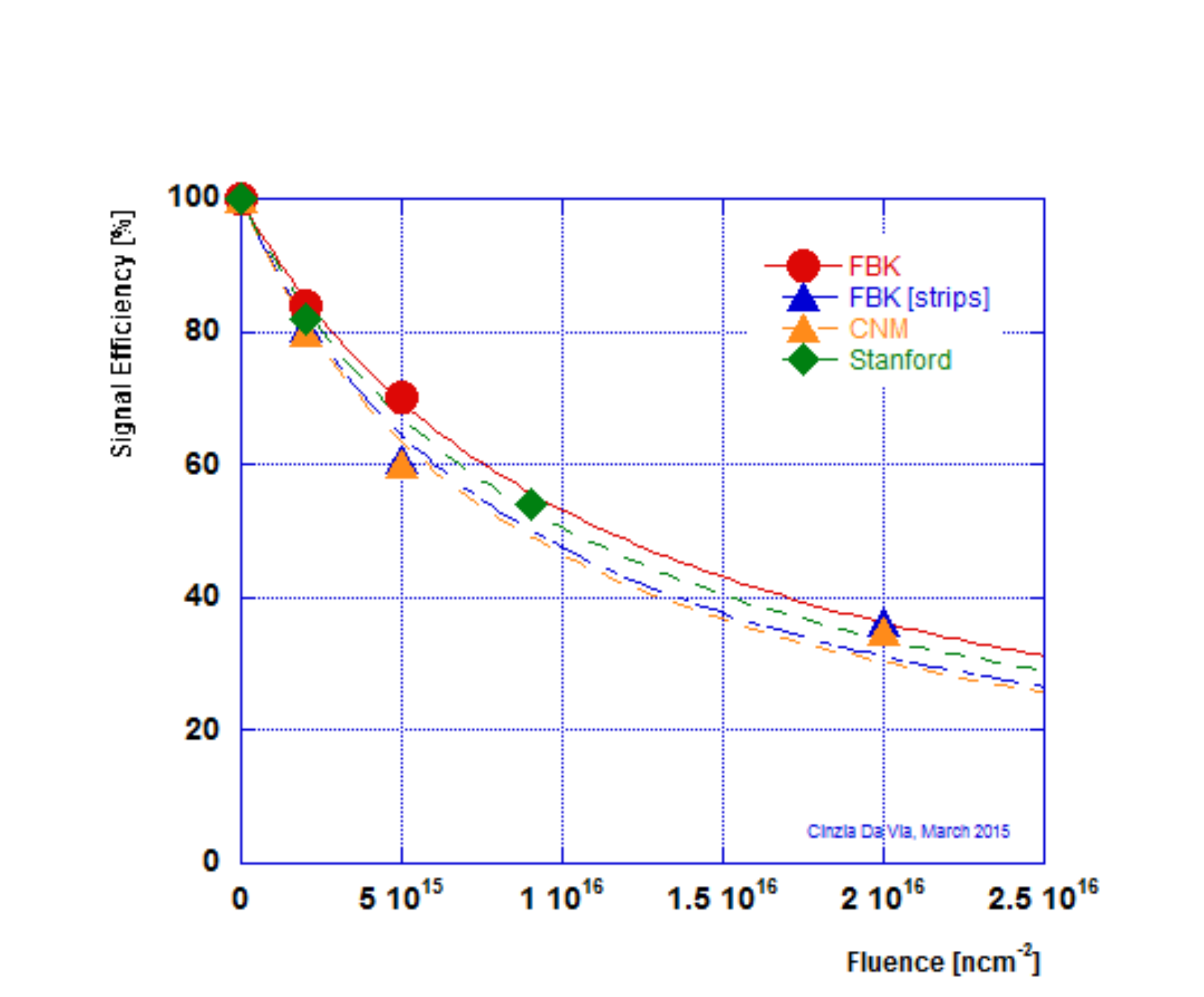}
	\caption{Signal efficiency as a function of fluence for FBK pixels~\cite{bib:IBLprototypes} (thickness d=230\,$\mu$m, inter-electrode distance l=67\,$\mu$m), FBK strips~\cite{GFradiationHard} (d=230\,$\mu$m, l=56\,$\mu$m), CNM strips~\cite{Koehler}(d=285\,$\mu$m, l=56\,$\mu$m) and Stanford pad diodes~\cite{bib:Cinzia2009} (d=210\,$\mu$m, l=71\,$\mu$m). Compilation by~\cite{bib:Cinzia2013,GFreview}.}
	\label{fig:charge}
\end{figure}

The charge collection after irradiation and its efficiency in 3D detectors has been measured up to fluences of $2\times10^{16}$\,n$_{eq}$/cm$^2$, which are relevant for the HL-LHC tracker upgrades. Fig.~\ref{fig:charge} shows a compilation of the signal efficiency, defined as the charge collected normalised to the one before irradiation, as a function of fluence for different 3D devices with inter-electrode distances between 56 and 71\,$\mu$m. At the IBL reference fluence of $5\times10^{15}$\,n$_{eq}$/cm$^2$, signal efficiencies of 60--70\% are achieved at moderate voltages of about 160\,V, whereas at the HL-LHC fluence of $2\times10^{16}$\,n$_{eq}$/cm$^2$ still more than 30\% is achieved at only 200\,V. It should be noted that the signal efficiency in 3D detectors significantly increases for lower inter-electrode distance. Comparisons of pixel detectors with varying electrode distance have confirmed this: signal efficiencies of 66, 51 and 36\% have been observed for 56, 71 and 103\,$\mu$m inter-electrode distance, respectively, measured at the same fluence of $9\times10^{15}$\,n$_{eq}$/cm$^2$~\cite{bib:Cinzia2009}. Hence, for future developments like HL-LHC 3D detectors, the signal efficiency is expected to be even further improved by reducing the electrode distance (see Sec.~\ref{sec:HLLHC}). The phenomenon of charge multiplication in highly irradiated detectors has been also observed in 3D detectors~\cite{Koehler} and can further enhance the signal after heavy irradiation at voltages above 200\,V, but at the cost of higher leakage current and noise.

\section{3D Detectors in the ATLAS Insertable B-Layer}
\label{sec:IBL}
In the first long shutdown of the LHC 2013-2015, the ATLAS experiment installed a new inner pixel layer called IBL at 3.3\,cm radius from the beam to enhance the vertex performance and robustness of the tracking system~\cite{bib:IBL}. 3D pixel detectors constitute 25\% of the IBL, namely the outer part of the IBL barrel, whereas traditional planar n-in-n sensors of 200\,$\mu$m thickness are used in the central part (75\%). Hence, since the start-up of the LHC Run 2 in June 2015, 3D detectors are for the first time part of a running HEP experiment and record collision data, which is considered a milestone for the technology.

Initially, 3D sensors from all above-mentioned vendors have been considered~\cite{bib:IBL3Dprod}. Due to their higher maturity and better yields, the final IBL sensors were produced by FBK~\cite{bib:FBKIBLProduction} and CNM~\cite{bib:CNMIBLProduction} in the double-sided process with the technology differences explained in Sec.~\ref{sec:technology}. IBL pixel sensors match the geometry of the FE-I4 readout chip~\cite{bib:FEI4} with a matrix of 80$\times$336 pixels of 250$\times$50\,$\mu$m$^2$ size and hence a total active area of 2.00$\times$1.68\,cm$^2$. Each 3D pixel consists of two n$^+$ junction columns surrounded by 6 p$^+$ ohmic columns that are shared with the neighbours as shown in Fig.~\ref{fig:FBKCNMdesign} (bottom), resulting in an inter-electrode distance of 67\,$\mu$m. The substrate is 230\,$\mu$m thick p-type silicon. Along the columns, slim edges of only 200\,$\mu$m width have been implemented to abut the sensors along the IBL barrel axis with minimal loss of sensitive area.

The IBL 3D project has been crucial to prove the maturity of the technology and gain important experience and insights, both from the technology and the physics-performance point of view. In particular it has been very instructive that during the sensor qualification and production, planar and 3D detectors of different technologies have been compared under similar conditions and with the same requirements~\cite{bib:IBLprototypes}. 

\begin{figure}[bt]
	\centering
	 \includegraphics[width=10cm]{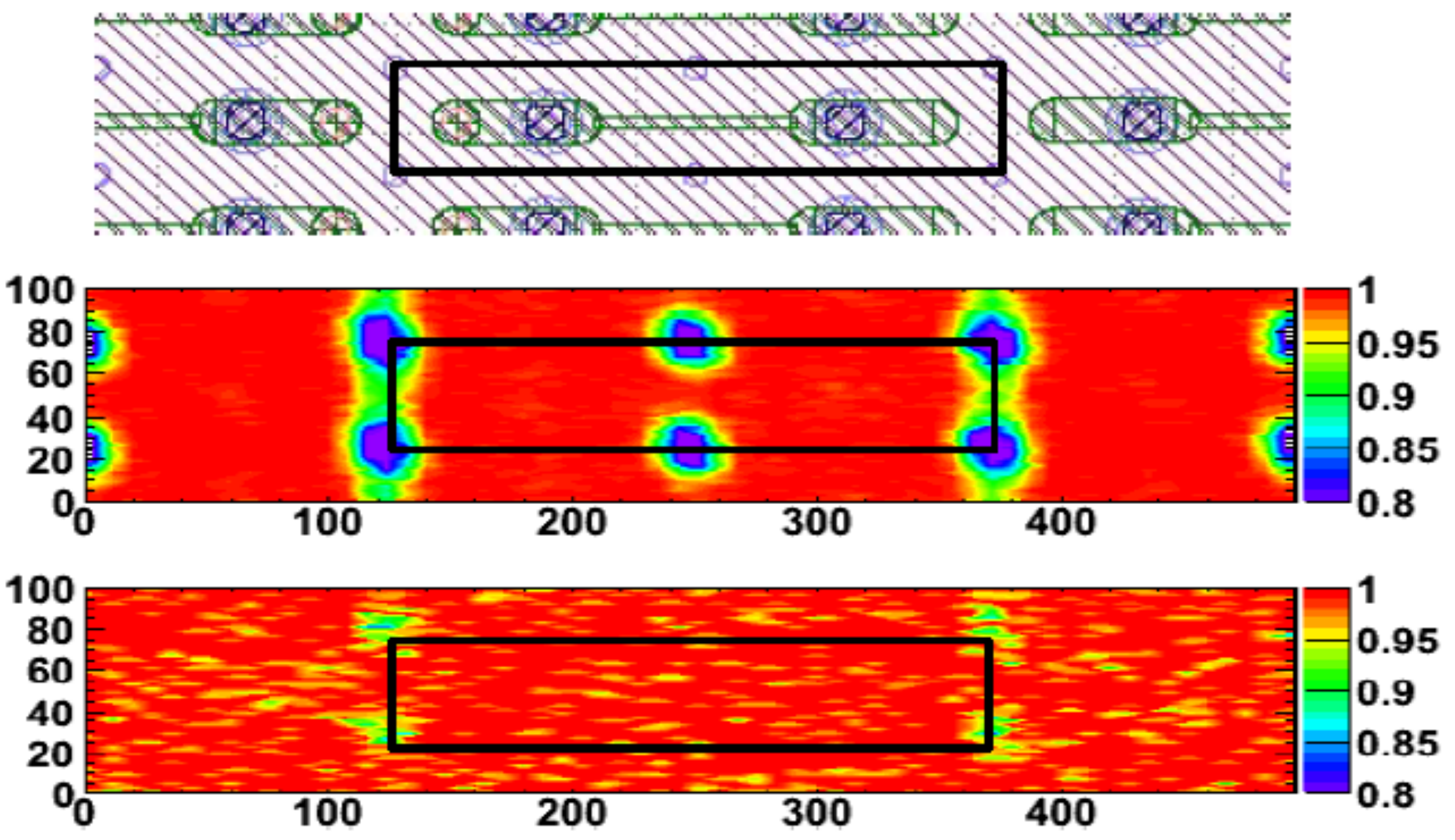}
	\caption{Hit efficiency maps of CNM FE-I4 pixel detectors corresponding to the 2$\times$2 pixel area shown on top. The devices are irradiated to the IBL target fluence of $5\times10^{15}$\,n$_{eq}$/cm$^2$, biased at 160\,V and oriented at perpendicular beam incidence (centre) and with a tilt of 15$^\circ$ (bottom) with average efficiencies of 97.5\% and 99.0\%, respectively~\cite{bib:IBLprototypes}.}
	\label{fig:IBLeff}
\end{figure}

It has been shown that a medium-scale production of 3D sensors is feasible already today with a reasonable production yield of 50--60\%~\cite{bib:IBL3Dprod}. It turned out that the sensor-selection method based on temporary metals (FBK) is a good indicator for the leakage currents of the assembled devices, whereas the guard-ring method (CNM) provides only a poor correlation~\cite{NanniIBL}. The breakdown voltages of 3D devices of about 20--100\,V were observed to be lower than for planar ones with more than 300\,V, but for 3D detectors much lower operational voltages of about 20\,V before irradiation are sufficient compared to about 80\,V for planar ones~\cite{NanniIBL}. FBK devices were observed to tend to lower breakdown voltages than CNM ones, which is understood as mainly arising from higher electric fields around the contact between the end of the through-passing junction column and the p-spray at the back side~\cite{FBKBreakdown,GFreview}. Due to the larger capacitance of 3D sensors (170\,fF) compared to planar ones (110\,fF), the noise of 3D sensors is a bit larger (130--140\,e$^-$ vs. 110\,e$^-$), with FBK sensors having slightly larger noise than CNM ones due to the larger overlap between the junction and ohmic columns~\cite{NanniIBL}. FBK sensors have been observed to reach the plateau of full charge collection at lower voltages than CNM devices due to an earlier depletion of the whole volume owing to the through-passing columns~\cite{bib:IBLprototypes}. A major focus of the qualification was the quest for high radiation hardness to cope with the IBL requirements of sustaining a fluence of $5\times10^{15}$\,n$_{eq}$/cm$^2$ expected for the whole life time equivalent to 500\,fb$^{-1}$. Already in Sec.~\ref{sec:CC} the high signal efficiency of 60--70\% for the IBL target fluence was mentioned. Test-beam measurements have confirmed a high average hit efficiency of >97\% at moderate voltages of only 160\,V for all devices measured (some of them even at lower voltages such as 120\,V), even at perpendicular beam incidence with some residual low-efficiency regions from the dead material of the 3D columns (see Fig.~\ref{fig:IBLeff} centre)~\cite{bib:IBLprototypes}. However, a more uniform response could be recovered with a tilt of 15$^{\circ}$, increasing the average efficiency by one percent (Fig.~\ref{fig:IBLeff} bottom). The power dissipation at 160\,V was measured to be about 10--15\,mW/cm$^2$ at the operating temperature of -15$^{\circ}$C. Hence, in comparison to the planar IBL sensors, which achieved similar hit efficiencies at voltages close to 1000\,V with power dissipations of about 90\,mW/cm$^2$, 3D sensors demonstrated superior radiation hardness in the sense that much lower operation voltages and cooling requirements are needed for such high fluences.

\section{3D Detectors in Forward Detectors (AFP and PPS)}
\label{sec:Forward}
Based on their high radiation hardness, the slim inactive edges and the maturity proven by the IBL production, 3D detectors have been chosen for further applications in future forward trackers of ATLAS and CMS-TOTEM, namely the ATLAS Forward Proton (AFP) detector~\cite{AFPTDR,AFP3D} and the Precision Proton Spectrometer (PPS)~\cite{PPSTDR}. Both are intended to tag very forward protons emerging intact from the proton-proton interaction point (IP) and will be placed in two Roman Pots at each side of the IP only a few\,mm away from the beam line at about 200\,m away from the collisions. The installations are targeted already for the end-of-the-year shutdown 2015/2016 (AFP) or next year (PPS).

\begin{figure}[bt]
	\centering
	 \includegraphics[width=15cm]{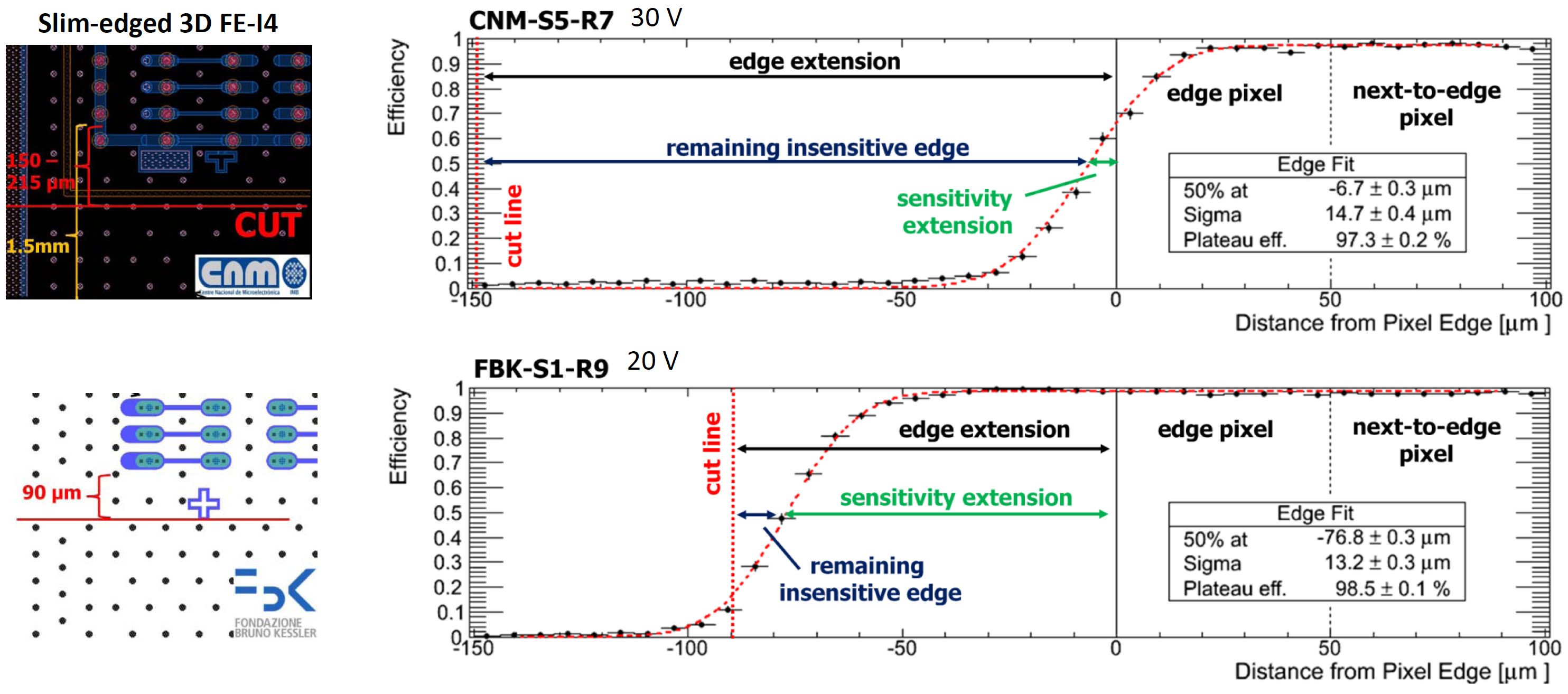}
	\caption{Left: Layout of CNM (top) and FBK (bottom) slim-edge AFP FE-I4 pixel sensor with cut line. Right: Corresponding hit efficiency around the edge pixels~\cite{AFP3D}. The smearing of the step is due to the telescope pointing resolution of about 15\,$\mu$m.}
	\label{fig:AFPslimEdge}
\end{figure}

Next to a good position resolution (10 and 30\,$\mu$m in horizontal and vertical direction, respectively), the requirements are mainly driven by the proximity of the detectors to the beam. This results in 1) a highly non-uniform fluence with an expected maximum fluence similar to IBL but orders of magnitude less only a few mm away from that maximum within the same sensor, and 2) the need for very slim inactive edges below 200\,$\mu$m for the sensor side facing the beam. AFP chose to base their detectors on the IBL FE-I4 sensors with the modification of a slim-edge cut line along the first pixel row opposite the wirebonds (see Fig.~\ref{fig:AFPslimEdge} left). PPS will use pixels based on the CMS PSI46dig readout chip~\cite{bib:PSI46dig} with 100$\times$150\,$\mu$m$^2$ pixels and similar slim edges. For AFP, a single FE-I4 device provides enough geometrical acceptance of slightly less than 4\,cm$^{2}$, whereas PPS uses a 6-chip module to achieve a similar area. AFP and PPS plan to install their modules in Roman Pot stations of 4 and 6 successive tracker planes with tilts of 13 and 20$^{\circ}$, respectively, in a telescope-like configuration.

During the AFP-sensor qualification, the scribe-cleave-passivate dicing technique~\cite{bib:AFP3D1} as well as a simple diamond-saw cut~\cite{AFP3D} was studied to obtain 100-200\,$\mu$m slim edges on FBK and CNM IBL sensors. Both achieved good results without large adverse effects on the break-down, noise and hit efficiencies. Hence the diamond-saw cut was chosen due to its simplicity. In beam tests dedicated to measure the hit efficiency around the slim edge, it was found that FBK and CNM sensors behaved differently due to the specific edge terminations. Fig.~\ref{fig:AFPslimEdge} (right) shows that CNM AFP prototypes are fully efficient up to the last pixel, from where on the 3D guard ring prevents charge collection from the area outside the pixelated region, resulting in about 150\,$\mu$m slim inactive edges. On the contrary, the FBK AFP prototypes also collect charge up to about 75\,$\mu$m away from the last pixel due to the absence of a guard ring, relying only on the ohmic guard fence for edge termination. The resulting <15\,$\mu$m insensitive edge is the slimmest edge to date apart from sensors with fully active edges. After irradiation, a decrease of the FBK sensitivity extension was observed as expected from trapping and the decrease of the depleted area, but at typical operation voltages of 150\,V, still 2/3 of the pre-irradiated value was maintained after $5\times10^{15}$\,n$_{eq}$/cm$^2$. Both CNM and FBK sensors fulfill the AFP requirements. Also PPS studied the slim-edge properties of FBK PSI46 3D sensors and found a sensitivity extension of 50 (70)\,$\mu$m for a tilt of 0 (20)$^{\circ}$, resulting in 150 (130)\,$\mu$m slim edges~\cite{PPSTDR,bib:PPS2}. The differences between AFP and PPS can be explained by the different pixel geometries and cut lines.

Concerning radiation hardness, AFP profits from the experience gained during the IBL qualification that had already proven the radiation hardness of FBK and CNM FE-I4 3D devices for uniform irradiation~\cite{bib:IBLprototypes} (see Sec.~\ref{sec:IBL}). Hence, dedicated AFP studies focused especially on non-uniform irradiation, which was achieved at CERN-PS with a focused 23\,GeV proton beam of 12\,mm FWHM and at Karlsruhe with holes in Aluminium masks for an even more localised irradiation with 23\,MeV protons~\cite{bib:AFP3D1,AFP3D}. The maxima of the fluence profiles were about $4\times10^{15}$\,n$_{eq}$/cm$^2$. In beam tests, hit efficiencies of 96--99\% were demonstrated in all regions of the non-uniformly irradiated device despite the locally very different fluences acquired, proving the capability of the FE-I4 3D devices to be operated under AFP conditions. Uniform irradiation studies were performed for CMS/PPS PSI46 3D devices. Previous investigations~\cite{bib:CMS1,bib:CMS2} were limited by the old version of the analog PSI46 readout chip that was not radiation hard enough, but a preliminary analysis of recent studies with the new digital PSI46dig chip could demonstrate hit efficiencies of 98\% (93\%) after 1 (3) $\times10^{15}$\,n$_{eq}$/cm$^2$ with FBK 3D sensors with only one 3D junction column per pixel~\cite{bib:PPS2}. The performance is expected to improve with devices that include two junction columns per pixel.

Based on the successful qualifications, AFP and PPS are currently producing the 3D pixel modules for their trackers. For AFP, 
a sensor production run at CNM finished in July 2014 with 5 wafers of 8 FE-I4 sensors each (8 more wafers were lost due to machine malfunctions during the production). The dicing was performed at CNM to obtain slim edges of about 180\,$\mu$m. The yield turned out to be relatively low due to etching problems during the DRIE process, resulting in 14 sensors with breakdown voltages above 10\,V. For the first AFP installation step at the end of 2015, 8 good modules are needed, so that a good assembly yield will be crucial. The bump bonding and module assembly is on-going at IFAE (Barcelona)\footnote{Institut de F\'{i}sica d'Altes Energies (IFAE), 08193 Bellaterra (Barcelona), Spain}. To provide more good sensors for the completion of AFP next year and further upgrades, CNM has started a new production of AFP sensors in February 2015 after solving the yield issue. For PPS, PSI46dig 3D sensors were foreseen on the first 6" production of FBK~\cite{FBK6inch}. However, that 6" commissioning run turned out to have a low yield on large sensors due to local defects. Another production run for PPS is on-going at CNM with mostly two (but also one) junction columns per pixel and up to 6-chip sensors. For the first time, CNM is adapting the edge termination without a 3D guard ring in order to profit from the sensitivity extension in case of a guard-fence-only termination. The installation of the PPS 3D modules is foreseen for 2016.

\section{Development of 3D Detectors for the High-Luminosity Upgrades}
\label{sec:HLLHC}
The HL-LHC (also called phase 2) upgrade~\cite{bib:HLLHC} foreseen for 2024 up to peak luminosities in the order of $10^{35}$\,s$^{-1}$cm$^{-2}$ presents a challenge for the silicon tracking detectors in ATLAS and CMS due to highly increased occupancies and radiation levels. For the innermost pixel layer, fluences up to $2\times10^{16}$\,n$_{eq}$/cm$^2$ are expected after 3000\,fb$^{-1}$. To cope with the higher particle densities, it is planned to increase the granularity using reduced pixel sizes of 50$\times$50\,$\mu$m$^2$ or 25$\times$100\,$\mu$m$^2$. A radiation-hard readout chip with these pixel dimensions is under development by the RD53 collaboration~\cite{bib:RD53}. It will provide a lower noise and threshold (about 1000\,e$^-$) than previous chips. It requires a sensor capacitance (leakage current) less than 100\,fF (10\,nA) per pixel.

For the sensors, 3D detectors are promising candidates for the inner pixel layers due to the proven radiation hardness as demonstrated above. However, clearly the development of a new generation of sensors is needed to meet the stringent HL-LHC and chip requirements. This is under way and the main trends will be outlined below. 

Until the new 3D productions are finished, the limits of the existing 3D generation are being explored simultaneously. FE-I3 and FE-I4 pixel detectors of the IBL/AFP generation have been irradiated to HL-LHC fluences and laboratory and beam test studies are on-going. Another interesting investigation focuses on the behaviour of 50\,$\mu$m-pitch detectors at high-incidence angles (relevant for high pseudo-rapidity $|\eta|$ at HL-LHC detectors) since for particles traversing almost parallel to the surface, the deposited charge per pixel is dominated by the pixel size, not by the thickness as for perpendicular incidence. Hence, for 50\,$\mu$m pitch, only about 3.5\,ke$^-$ charge deposited per pixel are expected even before irradiation. The concern is not so much about the overall hit efficiency, which should be still high since at large angles many pixels will fire (\eg the cluster size at 80$^{\circ}$, corresponding to $|\eta|=2.4$, is 27 for 230\,$\mu$m thickness). However, inefficiencies per pixel can lead to cluster splitting if they happen in the middle of a large cluster and to a degradation of the resolution if the first or the last pixels of the cluster are not firing efficiently. First test-beam studies with 80$^{\circ}$ incidence angle with respect to the short 50\,$\mu$m side of FE-I4 3D pixel detectors and a threshold tuned to 1000\,e$^-$ show a good per-pixel efficiency above 99\% before irradiation~\cite{bib:HighAngle}. The analysis for devices irradiated up to $5\times10^{15}$\,n$_{eq}$/cm$^2$ is still on-going, but preliminary results indicate an efficiency above 80\% per pixel.

\begin{figure}[bt]
	\centering
	 \includegraphics[width=8cm]{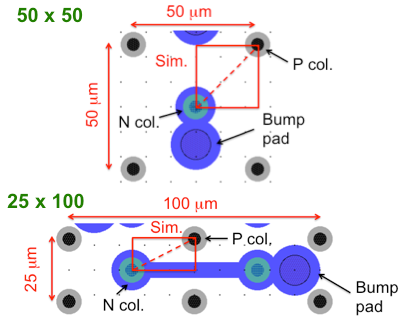}
	\caption{Layout of a one-junction-electrode 50$\times$50\,$\mu$m$^2$ pixel (top) and a two-junction-electrode 25$\times$100\,$\mu$m$^2$ pixel (bottom). Courtesy of G.-F. Dalla Betta~\cite{GFreview}. }
	\label{fig:HL-LHC3Dpixels}
\end{figure}

New technology developments for HL-LHC 3D detectors are pursued by all 3D vendors (Stanford, SINTEF, FBK, CNM). First of all, pixels with the new reduced cell sizes have to be successfully produced. Fig.~\ref{fig:HL-LHC3Dpixels} shows examples for a one-junction-electrode 50$\times$50\,$\mu$m$^2$ and a two-junction-electrode 25$\times$100\,$\mu$m$^2$ design with inter-electrode distances of 35 and 28\,$\mu$m, respectively. However, until the first RD53 chips will be available, the small pixel cells have to be matched to the existing readout chips with larger cell sizes. This can be done by shortening several pixels together to one large cell (at the cost of an increased capacitance) or by reading out only a fraction of the small pixels and shortening all other cells together and setting them to ground potential without readout~\cite{CNMITK}. 

It should be noted that the reduced pixel size automatically provides lower inter-electrode distances (by about a factor of 2 compared to the 67\,$\mu$m of IBL). Hence, improved radiation hardness is expected for the new generation as discussed Sec.~\ref{sec:CC}. However, a reduced electrode distance also implies a larger capacitance and, to keep the fraction of dead material constant for the higher column density, the column diameter has to be reduced from about 10\,$\mu$m today to typically 5\,$\mu$m for the new productions. To achieve this, one can either improve the aspect ratio (column thickness over diameter) from today's 20:1 to 40:1 for a constant sensor thickness. This is planned to be carried out by CNM using a cryogenic technique~\cite{3Detch,CNMITK}. Another option is to produce thinner sensors without the need to change the aspect ratio, as this will automatically result in narrower columns.

\begin{figure}[bt]
	\centering
	 \includegraphics[width=15cm]{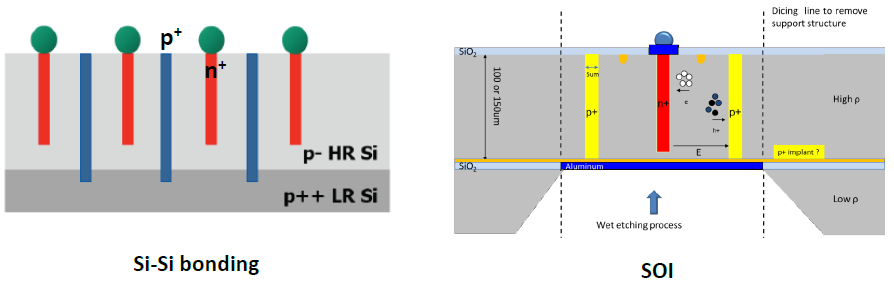}
	\caption{Layout of thin 3D sensors produced using the Si-Si wafer bonding (left) and the SOI (right) technique. Courtesy of G.-F. Dalla Betta and G. Pellegrini. }
	\label{fig:thin3D}
\end{figure}

The trend to produce thinner sensors, which is pursued by all vendors, has also other reasons besides the reduction of the 3D column diameter. It will reduce the sensor pixel capacitance, the leakage current, the cluster size at high incidence angles (\ie high $|\eta|$) for a better two-particle separation in a dense environment, as well as the multiple scattering for a better vertex resolution. A drawback is less deposited charge at low incidence angles (which is hoped to be compensated by a lower threshold and the smaller electrode distance leading to less trapping), and higher production and assembly complexity. CNM is trying to keep its double-sided technology, which is, however, expected to work only down to 200\,$\mu$m~\cite{CNMITK}. At the same time, all vendors, including CNM and FBK, are pursuing single-sided processes with a support wafer, which is necessary to advance to smaller thicknesses. FBK and Stanford are utilising the Si-Si wafer-bonding technique in which the thin sensitive high-resistivity substrate (100--130\,$\mu$m for FBK, 75--150\,$\mu$m for Stanford) is bonded to a low-resistivity support wafer~\cite{GFreview,FBKsingleSide}, whereas CNM (100--150\,$\mu$m) and SINTEF (50--100\,$\mu$m) produce their sensors on silicon-on-insulator (SOI) wafers~\cite{CNMITK} (see Fig.~\ref{fig:thin3D}). In all cases, both column types are etched from the front side, but the junction columns stop shortly before reaching the support wafer, whereas the ohmic columns pass through. This allows, after thinning the support wafer, to apply the bias voltage to the ohmic columns from the back side, thereby facilitating the assembly process compared to previous single-sided approaches with a bias tab. 

Further trends in 3D technology development include the following (incomplete list), which is partly based on the experience gained in previous productions as described above. FBK and SINTEF will produce on 6" wafers to reduce costs~\cite{FBK6inch}. CNM is implementing new on-wafer selection methods like poly-Si resistors to replace the guard ring method~\cite{CNMITK}. FBK optimised their process, using now non-passing-through junction columns to improve the breakdown behaviour of their sensors~\cite{FBKBreakdown,GFreview}. Stanford is trying the implementation of junction columns of varying thickness in the same substrate for higher flexibility of optimising the sensitive thickness, as well as for possibly improved position resolution and two-particle separation at high incidence angles if the varying column thickness is implemented even in one sensor~\cite{varyingThickness}. And all vendors are implementing and optimising either active or slim edges  (\eg as already for the PPS production, CNM is now also producing part of the sensors without guard ring~\cite{CNMITK}).

\section{Conclusions}

State-of-the art 3D silicon sensors present many advantages such as superior radiation hardness and slim insensitive edges. The technology has proven maturity with the first medium-scale production for the ATLAS IBL, where 3D pixel detectors take data since June 2015 for the first time in a HEP experiment. The second use in forward detectors like AFP and PPS is imminent after a successful qualification proving the feasibility of very slim edges of 10--150\,$\mu$m and the operability after non-uniform irradiation. Module production and assembly is on-going. 3D sensors are also promising candidates for the innermost pixel layers of the future HL-LHC trackers where they need to cope with harsh radiation environments and occupancies. Present-day devices are being evaluated up to HL-LHC fluences and new technology developments are being pursued to address the challenges of ultra radiation hardness, smaller cell sizes with narrower columns, thinner devices, better breakdown behaviour, improved on-wafer selection and lower costs. First results are expected in the course of next year, in time for being considered for the design of the HL-LHC pixel detectors.

\acknowledgments

The author wishes to thank S.\,Grinstein, I.\,Lopez Paz, E.\,Cavallaro, G.\,Pellegrini, G.-F.\,Dalla Betta, C.\,Da Via, F.\,Ravera and the whole ATLAS, CMS, RD50, CNM and FBK 3D communities for fruitful discussions and help with the material. This work was partially funded by the MINECO, Spanish Government, under grants FPA2013-48308 and SEV-2012-0234 (Severo Ochoa excellence program) and the European Commission under the FP7 Research Infrastructures project AIDA, grant agreement no. 262025.


\begin{thebibliography}{99}

\bibitem{Parker3D} 
S.I. Parker, C.J. Kenney and J. Segal, 
\emph{3D - A proposed new architecture for solid-state silicon detectors}, 
\href{http://dx.doi.org/10.1016/S0168-9002(97)00694-3}
{\emph{Nucl. Instrum. Meth.} \textbf{A 395} (1997) 328}.

\bibitem{bib:Cinzia2009}
C. Da Via et al., 
\emph{3D active edge silicon sensors with different electrode configurations: Radiation hardness and noise performance}, 
\href{http://dx.doi.org/10.1016/j.nima.2009.03.049}
{\emph{Nucl. Instrum. Meth.} \textbf{A 604} (2009) 505}.

\bibitem{Koehler}
M. K{\"o}hler et al., 
\emph{Comparative measurements of highly irradiated n-in-p and p-in-n 3D silicon strip detectors}, 
\href{http://dx.doi.org/10.1016/j.nima.2011.08.041}
{\emph{Nucl. Instrum. Meth.} \textbf{A 659} (2011) 272}.

\bibitem{GFradiationHard}
G.-F. Dalla Betta et al., 
\emph{Radiation hardness tests of double-sided 3D strip sensors with passing-through columns}, 
\href{http://dx.doi.org/10.1016/j.nima.2014.05.007}
{\emph{Nucl. Instrum. Meth} \textbf{A 765} (2014) 155}.

\bibitem{bib:activeEdge1}
C.J. Kenney, S. Parker and E. Walckiers,
\emph{Results from 3-D silicon sensors with wall electrodes: near-cell-edge sensitivity measurements as a preview of active-edge sensors},
\href{http://dx.doi.org/10.1109/23.983250}
{\emph{IEEE Trans. Nucl. Sci.} \textbf{48} (2001) 2405}.

\bibitem{FBKcutStudy1}
M. Povoli et al., 
\emph{Slim edges in double-sided silicon 3D detectors}, 
\href{http://dx.doi.org/10.1088/1748-0221/7/01/C01015}
{\emph{JINST} \textbf{7} (2012) C01015}.

\bibitem{FBKcutStudy2}
M. Povoli et al.,
\emph{Design and testing of an innovative slim-edge termination for silicon radiation detectors}, 
\href{http://dx.doi.org/10.1088/1748-0221/8/11/C11022}
{\emph{JINST} \textbf{8} (2013) C11022}.

\bibitem{AFP3D}
J. Lange, E. Cavallaro, S. Grinstein and I. Lopez Paz,
\emph{3D silicon pixel detectors for the ATLAS Forward Physics experiment},
\href{http://dx.doi.org/10.1088/1748-0221/10/03/C03031}
{\emph{JINST} \textbf{10} (2015) C03031}.

\bibitem{ATLAS} \AC, 
\emph{The ATLAS experiment at the CERN Large Hadron Collider},
\href{http://iopscience.iop.org/1748-0221/3/08/S08003}
{\emph{JINST} \textbf{3} (2008) S08003}.


\bibitem{bib:IBL}
ATLAS Collaboration,
\emph{ATLAS Insertable B-Layer Technical Design Report},
\href{https://cds.cern.ch/record/1291633}
{CERN-LHCC-2010-013, ATLAS-TDR-19 (2010)}.

\bibitem{bib:IBL3Dprod}
C. Da Via et al.,
\emph{3D silicon sensors: Design, large area production and quality assurance for the ATLAS IBL pixel detector upgrade},
\href{http://dx.doi.org/10.1016/j.nima.2012.07.058}
{\emph{Nucl. Instrum. Meth.} \textbf{A 694} (2012) 321}.

\bibitem{bib:IBLprototypes}
ATLAS IBL Collaboration, 
\emph{Prototype ATLAS IBL modules using the FE-I4A front-end readout chip}, 
\href{http://dx.doi.org/10.1088/1748-0221/7/11/P11010}
{\emph{JINST} \textbf{7} (2012) P11010}.

\bibitem{AFPTDR}
ATLAS Collaboration,
\emph{Technical Design Report for the ATLAS Forward Proton Detector},
\href{https://cds.cern.ch/record/2017378}
{CERN-LHCC-2015-009, ATLAS-TDR-024 (2015)}.

\bibitem{CMS} \CC, 
\emph{The CMS experiment at the CERN LHC},
\href{http://iopscience.iop.org/1748-0221/3/08/S08004}
{\emph{JINST} \textbf{3} (2008) S08004}.

\bibitem{TOTEM} TOTEM Collaboration, 
\emph{The TOTEM experiment at the CERN Large Hadron Collider}
\href{http://iopscience.iop.org/1748-0221/3/08/S08007}
{\emph{JINST} \textbf{3} (2008) S08007}.


\bibitem{PPSTDR}
CMS-TOTEM Collaboration,
\emph{CMS-TOTEM Precision Proton Spectrometer},
\href{https://cds.cern.ch/record/1753795}
{CERN-LHCC-2014-021, TOTEM-TDR-003, CMS-TDR-13 (2014)}.

\bibitem{bib:HLLHC}
F. Gianotti et al.,
\emph{Physics potential and experimental changes of the LHC luminosity upgrade},
\href{http://refhub.elsevier.com/S0168-9002(14)00506-3/sbref3}
{\emph{Eur. Phys. J.} \textbf{C 39} (2005) 293}.


\bibitem{3Detch}
B. Wu et al., 
\emph{High aspect ratio silicon etch: A review}, 
\href{http://dx.doi.org/10.1063/1.3474652}
{\emph{J. Appl. Phys.} \textbf{108} (2010) 051101 }.

\bibitem{SNF}
C.J. Kenney, S.I. Parker, J. Segal and C. Storment,
\emph{Silicon detectors with 3-D electrode arrays: fabrication and initial test results},
\href{http://dx.doi.org/10.1109/23.785737}
{\emph{IEEE Trans. Nucl. Sci.}, \textbf{46(4)} (1999) 1224}.

\bibitem{sintef}
T.-E. Hansen et al., 
\emph{First fabrication of full 3D-detectors at SINTEF}, 
\href{http://dx.doi.org/10.1088/1748-0221/4/03/P03010}
{\emph{JINST} \textbf{4} (2009) P03010}.

\bibitem{CNM1}
G. Pellegrini et al.,
\emph{First double sided 3-D detectors fabricated at CNM-IMB},
\href{http://dx.doi.org/10.1016/j.nima.2008.03.119}
{\emph{Nucl. Instrum. Meth.} \textbf{A 592} (2008) 38}.

\bibitem{bib:CNMIBLProduction}
G. Pellegrini et al.,
\emph{3D double sided detector fabrication at IMB-CNM},
\href{http://dx.doi.org/10.1016/j.nima.2012.05.087}
{\emph{Nucl. Instrum. Meth.} \textbf{A 699} (2013) 27}.

\bibitem{FBK1}
A. Zoboli et al.,
\emph{Double-Sided, Double-Type-Column 3D detectors at FBK: Design, Fabrication and Technology Evaluation},
\href{http://dx.doi.org/10.1109/TNS.2008.2002885}
{\emph{IEEE Trans. Nucl. Sci.} \textbf{NS-55(5)} (2008) 38}.

\bibitem{bib:FBKIBLProduction}
G. Giacomini et al.,
\emph{Development of double-sided full-passing-column 3D sensors at FBK},
\href{http://dx.doi.org/10.1109/TNS.2013.2262951}
{\emph{IEEE Trans. Nucl. Sci.} \textbf{NS-60(3)} (2013) 2357}.

\bibitem{bib:Cinzia2013}
C. Da Via et al., 
\emph{3D active edge silicon sensors: device processing, yield and QA for the ATLAS IBL production}, 
\href{http://dx.doi.org/10.1016/j.nima.2012.05.070}
{\emph{Nucl. Instrum. Meth} \textbf{A 699} (2013) 18}.

\bibitem{GFreview}
G.-F. Dalla Betta,
\emph{3D Silicon Detectors},
\href{http://pos.sissa.it/archive/conferences/219/013/IFD2014_013.pdf}
{\emph{PoS IFD2014 (2015) 013}}.

\bibitem{bib:FEI4}
M. Garcia-Sciveres et al.,
\emph{The FE-I4 pixel readout integrated circuit},
\href{http://dx.doi.org/10.1016/j.nima.2010.04.101}
{\emph{Nucl. Instrum. Meth.} \textbf{A 636} (2011) S155}.

\bibitem{NanniIBL}
G. Darbo on behalf of the ATLAS Collaboration,
\emph{Experience on 3D Silicon Sensors for ATLAS IBL},
\href{http://dx.doi.org/10.1088/1748-0221/10/05/C05001}
{\emph{JINST} \textbf{10} (2015) C05001}.

\bibitem{FBKBreakdown}
G.-F. Dalla Betta et al.,
\emph{Characterization of New FBK Double-Sided 3D Sensors with Improved Breakdown Voltage},
{\emph{Conference Record of 2013 IEEE Nuclear Science Symposium} N41-1}.

\bibitem{bib:PSI46dig}
\emph{The CMS Pixel Readout Chip for the Phase 1 Upgrade},
\href{http://stacks.iop.org/1748-0221/10/i=05/a=C05029}
{\emph{JINST} \textbf{10} (2015) C05029}.

\bibitem{bib:AFP3D1}
S. Grinstein et al.,
\emph{Beam test studies of 3D pixel sensors irradiated non-uniformly for the ATLAS forward physics detector},
\href{http://dx.doi.org/10.1016/j.nima.2013.03.064}
{\emph{Nucl. Instrum. Meth.} \textbf{A 730} (2013) 28}.

\bibitem{bib:PPS2}
F. Ravera et al.,
\emph{Results on FBK 3D pixel detectors for CMS},
\href{https://indico.cern.ch/event/351695/session/6/contribution/35}
{Presentation at: \emph{10th Anniversary Trento Workshop on Advanced Silicon Radiation Detectors}, Trento, Italy, 18 February 2015}.

\bibitem{bib:CMS1}
F. Mu{\~n}oz Sanchez,
\emph{Study of New Silicon Sensors for Experiments at Future Particle Colliders},
\href{http://jinst.sissa.it/jinst/theses/2014_JINST_TH_001.jsp}
{PhD thesis, Universidad de Cantabria (Santander), 2014}.

\bibitem{bib:CMS2}
A. Krzywda et al.,
\emph{Pre- and post-irradiation performance of FBK 3D silicon pixel detectors for CMS},
\href{http://dx.doi.org/10.1016/j.nima.2014.06.029}
{\emph{Nucl. Instrum. Meth.} \textbf{A 763} (2014) 404}.

\bibitem{FBK6inch}
M. Boscardin et al.,
\emph{Preliminary results from the first batch of Si 3D sensors fabricated at FBK on 6 inch wafers},
\href{https://indico.cern.ch/event/351695/session/6/contribution/42}
{Presentation at: \emph{10th Anniversary Trento Workshop on Advanced Silicon Radiation Detectors}, Trento, Italy, 18 February 2015}.

\bibitem{bib:RD53}
RD53 Collaboration,
\emph{RD Collaboration Proposal: Development of pixel readout integrated circuits for extreme rate and radiation},
\href{https://cds.cern.ch/record/1553467}
{CERN-LHCC-2013-008 (2013)}.

\bibitem{bib:HighAngle}
I. Lopez Paz et al.,
\emph{Recent testbeam results of 50\,$\mu$m pitch 3D sensors at high incidence angle for HL-LHC},
\href{https://indico.cern.ch/event/381195/session/1/contribution/20}
{Presentation at: \emph{26th RD50 Workshop}, Santander, Spain, 22 July 2015}.

\bibitem{CNMITK}
M. Baselga et al.,
\emph{New 3D fabrication at CNM for the ATLAS Pixel upgrade},
\href{https://indico.cern.ch/event/351695/session/9/contribution/15}
{Presentation at: \emph{10th Anniversary Trento Workshop on Advanced Silicon Radiation Detectors}, Trento, Italy, 19 February 2015}.

\bibitem{FBKsingleSide}
G.-F. Dalla Betta et al.,
\emph{Development of a new generation of 3D pixel sensors for HL-LHC},
\href{http://dx.doi.org/10.1016/j.nima.2015.08.032}
{\emph{Nucl. Instrum. Meth.} \textbf{A} (2015) in press}.


\bibitem{varyingThickness}
C. Da Via et al.,
\emph{3D silicon sensors with variable electrode depth for radiation hard high resolution particle tracking},
\href{http://dx.doi.org/10.1088/1748-0221/10/04/C04020}
{\emph{JINST} \textbf{10} (2015) C04020}.




\end{thebibliography}
\end{document}